# Laboratory simulations of the Vis-NIR spectra of comet 67P using sub-μm sized cosmochemical analogues

B. Rousseau[a,*], S. Érard[a], P. Beck[b], É. Quirico[b], B. Schmitt[b], O. Brissaud[b], G. Montes-Hernandez[c], F. Capaccioni[d], G. Filacchione[d], D. Bockelée-Morvan[a], C. Leyrat[a], M. Ciarniello[d], A. Raponi[d], D. Kappel[e], G. Arnold[e], L.V. Moroz[e,f], E. Palomba[d], F. Tosi[d], the VIRTIS Team

[a] LESIA, Observatoire de Paris, PSL Research University, CNRS, Sorbonne Universités, UPMC Univ. Paris 06, Univ. Paris Diderot, Sorbonne Paris Cité, 5 place Jules Janssen, 92195 Meudon, France
[b] UniversitéGrenoble Alpes, CNRS, Institut de Planétologie et d'Astrophysique de Grenoble (IPAG), UMR 5274, Grenoble F-38041, France
[c] UniversitéGrenoble Alpes, ISTerre, F-38041 Grenoble, France
[d] IAPS-INAF, Istituto di Astrofisica e Planetologia Spaziali, Area di Ricerca di Tor Vergata,Via del Fosso del Cavaliere, 100, 00133, Rome, Italy
[e] Institute for Planetary Research, German Aerospace Center (DLR), Rutherfordstraße 2, 12489 Berlin, Germany
[f] Institute of Earth and Environmental Science, University of Potsdam, Potsdam, Germany

## Abstract

Laboratory spectral measurements of relevant analogue materials were performed in the framework of the Rosetta mission in order to explain the surface spectral properties of comet 67P. Fine powders of coal, iron sulphides, silicates and their mixtures were prepared and their spectra measured in the Vis-IR range. These spectra are compared to a reference spectrum of 67P nucleus obtained with the VIRTIS/Rosetta instrument up to 2.7 μm, excluding the organics band centred at 3.2 μm. The species used are known to be chemical analogues for cometary materials which could be present at the surface of 67P. Grain sizes of the powders range from tens of nanometres to hundreds of micrometres. Some of the mixtures studied here actually reach the very low reflectance level observed by VIRTIS on 67P. The best match is provided by a mixture of sub-micron coal, pyrrhotite, and silicates. Grain sizes are in agreement with the sizes of the dust particles detected by the GIADA, MIDAS and COSIMA instruments on board Rosetta. The coal used in the experiment is responsible for the spectral slope in the visible and infrared ranges. Pyrrhotite, which is strongly absorbing, is responsible for the low albedo observed in the NIR. The darkest components dominate the spectra, especially within intimate mixtures. Depending on sample preparation, pyrrhotite can coat the coal and silicate aggregates. Such coating effects can affect the spectra as much as particle size. In contrast, silicates seem to play a minor role.

## 1. Introduction

After spending about 2 years in the vicinity of comet 67P, the Rosetta spacecraft has provided unprecedented amounts of information on the formation and evolution of comets and the surface processes at play during a revolution. Among the experiments aboard the Rosetta spacecraft, the Visible and Infrared Thermal Imaging Spectrometer (VIRTIS, Coradini et al., 2007) instrument was able to determine the spectral signature of cometary dust and its mixture with $CO_2$ and $H_2O$ ices. VIRTIS observations have revealed that the surface dust is generally homogeneous across the nucleus (with no obvious differences between the two lobes), is extremely dark (0.06 geometric albedo in the visible, (Ciarniello et al., 2015; Fornasier et al., 2015), comprised in the 0.04–0.07 range of other comets, Ciarniello et al. (2015)), and that the surface material is spectrally red across the 0.3–2.7 μm range. While the presence of a 3.2 μm absorption band suggests the presence of some organic components within the dust, significant efforts remain to be done to understand the compositional controls on the spectral signature of 67P and more generally comets and related D-type asteroids (Vernazza and Beck, 2016). Instruments onboard Rosetta have provided information on the physics of 67P dust in the coma, including dust size and velocity distribution. The Grain Impact Analyser and Dust Accumulator (GIADA, (Colangeli et al., 2007; Della Corte et al., 2014)) measured sizes from 100 μm to millimetres and characterised the dust flux (Rotundi et al., 2015; Della Corte et al., 2016; Fulle et al., 2016b). The Cometary Secondary Ion Mass Analyser (COSIMA, Henkel et al., 2003) measured sizes from tens of micrometres to millimetres (Langevin et al., 2016; Merouane et al., 2016). The Micro-Imaging Dust Analysis System (MIDAS, Riedler et al. (2007)), which had the capability to observe grain-size at the smallest scale within the Rosetta payload, was able to image particles from nanometres to micrometres (Bentley et al., 2016; Mannel et al., 2016). The measurements of these instruments indicate the presence of fluffy aggregates of several hundred micrometres (Langevin et al., 2016) which are composed of a sub-structure ranging from micrometre to sub-micrometre sizes (Bentley et al., 2016; Mannel et al., 2016). These fluffy particles are the parents of the individual grains detected by GIADA and "account for ≈15% of the total non-volatile volume, but for < 1% of the total non-volatile mass" (Fulle et al., 2016a; 2015). Finally, the latter results reveal that the building blocks of cometary dust are fine-grained, which was expected from thermal emission spectroscopy of cometary dust tails (Kolokolova et al., 2004) and coma (Bockelée-Morvan et al. (2017)). Of importance here is the fact that the major part of the building blocks are sub-µm, therefore of the order of the wavelength when observed in the Vis-NIR: a good cometary analogue requires the right grain size, in order to be in a realistic regime of radiative transfer. The findings summarized above help in the selection of cometary analogue materials, in terms of their particle size and their composition. Here, by means of laboratory measurements of the Vis-NIR reflectance properties of these materials and their mixtures, we aim to reproduce the spectral signature of 67P as measured by the VIRTIS instrument. In particular, our approach is to use sub-µm powders, analogues to cometary dust, and to build on the vast knowledge derived from laboratory studies of IDP in order to define the constituents and composition of our analogue material.

## 2. Methods

### 2.1. Compositional endmembers: rationale

The composition of the refractory component of 67P was inferred from measurements of the VIRTIS and COSIMA instruments aboard Rosetta (Capaccioni et al., 2015; Quirico et al., 2016; Fray et al., 2016), from ground and satellite observations (Crovisier et al., 1904; Sugita et al., 2005; Harker et al., 2005; Lisse et al., 2006), and from the composition of STARDUST grains, chondritic porous stratospheric IDPs and micrometeorites (Keller et al., 2006; Zolensky et al., 2006; Dobrica and Brearley, 2011; Dobrica et al., 2011; 2012). The presence of mafic minerals (olivine, pyroxene) and glasses has been evidenced by spectroscopic observations and laboratory analysis of cosmomaterials (Dobrica and Brearley, 2011; Dobrica et al., 2012; Engrand et al., 2016). Amorphous silicates are also determined by the analysis in the laboratory of presumed cometary grains and the most primitive chondrites, which contain the so-called GEMS (Glass with Embedded Metals and Sulphides) (Leroux et al., 2015). GEMS comprise a Mg-rich silicate glass phase that hosts very small inclusions of Fe metal and Fe sulphides. The contribution of these sub-µm Fe-rich opaques to the spectral signature of 67P has been discussed in Quirico et al. (2016) and the presence of sulphides was suggested in the spectral modelling of VIRTIS data (Capaccioni et al., 2015). Complex macromolecular organics have been observed in situ on Halley and 67P, and they are ubiquitous in presumed cometary grains as IDPs and micrometeorites (Fomenkova et al., 1994 ; Fray et al., 2016 ; and see discussion in Quirico et al., 2016). This organic component is abundant in stratospheric IDPs (15wt.% in average, up to 90wt.%) and a significant fraction of organics in 67P dust is needed to explain the dielectric properties of the cometary nucleus (Hérique et al., 2016). According to these results, three main groups of materials can be finally identified on cometary nuclei in addition to the volatiles ($H_2O$, $CO_2$, CO, …):

i. anhydrous silicates such as Mg-rich olivines, pyroxenes and glasses;
ii. opaque minerals such as sulphides (pyrrhotite $Fe_{1-x}S$ with $0 < x < 0.20$) and Ni-bearing iron $Fe_0$ ;
iii. a refractory polyaromatic carbonaceous material.

Below, the analogue materials selected for each group are reported.

**Group (i):** a terrestrial silicate was used as analogue (Table 1): a dunite from an Oman ophiolite rock, which contains 95% of olivine ($[Fe_{0.2}Mg_{0.8}]_4SiO_4$), and minor amounts of spinel and phlogopite ($KMg_3(Si_3Al)O_{10}(OH)_2$). This sample displays signatures of $H_2O/OH$ in its reflectance spectrum, due to the presence of small amounts of phyllosilicates as alteration products. We did not use glass samples, but we expect this component to behave as a bright phase due to the lack of iron. Finally, it is important to note that dunite may also contain minor amount of oxides and opaque phases.

Group (ii): A pyrrhotite sample (provided by Museum National d' Histoire Naturelle, Paris, France) was used as an analogue of opaque minerals. Its composition was estimated to be $Fe_{0.35}S$ by X-ray fluorescence. Pyrrhotite is sensitive to oxidation and was stored in a desiccator maintained under primary vacuum. Nevertheless, oxidation became very critical for ground samples containing submicrometre-sized grains. In one case, we observed an exothermic reaction during grinding that led to a brownish sample. X-ray diffraction measurements showed that this sample contained pyrrhotite 46%, szomolnokite ($FeSO_4 \cdot H_2O$) 38%, elemental sulfur 13% and magnetite 3%. In other cases, a black powder was recovered, with a reflectance level lower than that of the brownish sulphate-rich sample. No iron metal was used due to the critical instability of iron nanoparticles exposed to air, which immediately got oxidised. We consider here that the sulphide component mimics the overall behaviour of all opaque materials.

**Table 1** - Material endmembers used in the synthesis of cometary analogues.

| Sample | Species | Chemical composition |
| --- | --- | --- |
| Coal | Lignite (PSOC 1532) | $C_{70.62}H_{5.32}O_{23.12}N_{0.94}$ |
| Iron sulphide | Pyrrhotite | $Fe_{1-x}S$ with $0<x<0.20$ |
| Silicate | Dunite (95% olivine) | $[Fe_{0.2}Mg_{0.8}]_2SiO_4$ |

Group (iii): A subbituminous C (PSOC 1532, provided by the Penn State University Data Bank) was used as an analogue of cometary refractory carbonaceous material. This low maturity coal (huminite reflectance 0.33%) contains three main macerals (huminite 78.5%, inertinit 13.1% and liptinite 9.2%) that correspond to the main constituents of plants. These macerals display chemical differences, but after grinding they are intimately mixed. The texture of the initial plant precursors does not appear in these macerals as in peats. Nevertheless, coals are microporous due to the imperfect stacking of polyaromatic units that generates voids in the carbon skeleton. We do not have estimates of this microporosity, which is also counterbalanced by the presence of a soluble component (coals are described as gels). He has a composition fairly similar to chondritic IOM from primitive chondrites (Alexander et al., 2007), a disordered polyaromatic structure similar to that of refractory organics in stratospheric IDPs of presumed cometary origin (see discussion in Quirico et al., 2016). However, this coal does not strictly mimic an insoluble organic matter as isolated from primitive chondrites, but a mix of IOM and of semi-volatile solvent extractable component. The elemental composition of PSOC 1532 is $C_{70.62}H_{5.32}O_{23.12}N_{0.94}$, which is fairly similar to that of Insoluble Organic Matter extracted from primitive chondrites. This composition may be slightly different than that of cometary dust (Aléon et al., 2001; Fray et al., 2016), but the critical point here is to get a material with a polyaromatic structure, which mostly controls the optical properties (see discussion in Quirico et al., 2016).

### 2.2. Sample preparation and characterisation

The endmembers defined above (Table 1) were ground and sieved. Raw materials were first hand-crushed and ground in a mortar then sieved to sort out grains < 200 µm. Hand crushing was however not very suitable for soft and sticky coal samples, which usually led to large compact agglomerates. In this case, a dry 90 min grinding was conducted with a planetary grinder Retsch PM100, mounted with a $ZrO_2$ 50 ml bowl and 2 mm balls. The second step consisted in a colloidal (in ethanol) grinding of this size fraction using the same apparatus with 500 µm balls, a 550 rpm speed rotation, and stopping periods of 30 s every 10 min. Although the samples are heated up during the grinding, the temperature reached is not sufficient (< 50 °C) to affect their composition. The colloidal solution was then sieved again (< 25 µm) to remove the large clusters formed by re-agglomeration, or isolated big grains that might have escaped the grinding process and transferred into a desiccator (60 °C, primary vacuum) to remove ethanol. Finally, a crust composed of a very fine-grained material was recovered, which we hand-crushed to obtain free sub-micrometre-sized grains mixed up with agglomerates (Fig. 1). Binary (coal + pyrrhotite or coal + silicate) and ternary (coal + pyrrhotite + silicate) mixtures were prepared through two protocols. The first protocol was achieved with a MM200 grinder (Retsch). An agate mortar without milling balls was filled with several powders and shaken at 10-15 Hz for 5 min. The second protocol consisted in hand-mixing with a mortar and pestle. It was found much more efficient to reach fully intimate mixtures at a sub-micrometric scale (Fig. 2).

### 2.3. IR Reflectance spectroscopy

Reflectance measurements were performed with a spectrogonio-radiometer at the Institut de Planétologie et d'Astrophysique de Grenoble (IPAG). A full description of this instrument can be found in Bonnefoy (2001) and Brissaud et al. (2004). The instrument is made of two pivoting arms, one carrying the visible and infrared detectors, the other carrying the illumination source. Contrary to Bonnefoy (2001) and Brissaud et al. (2004), we used a new measurement mode of the spectro-gonio-radiometer. The initial concept (Brissaud et al., 2004) was to measure the reflected signal from a 2 cm diameter circular area within a 20 cm diameter illuminated area. In the new mode, the illumination is focused onto a 7.5 mm diameter area, without changing the size of the observed area. This new measurement mode provides an increased signal-to-noise ratio (x30), which is well adapted to studies of dark materials. Measurements were performed at 0° incidence, 30° emergence, 0° azimuth angles and under ambient atmosphere. The calibrated data, given in REFF (reflectance factor), were obtained by dividing the radiance signal coming from the detectors by the radiance from reference surfaces (Spectralon$^{TM}$-SR99 for the 0.4-2.5 μm range and Infragold$^{TM}$ for the 2.5-4.0 μm range). The spectral sampling of the measurements was 20 nm.

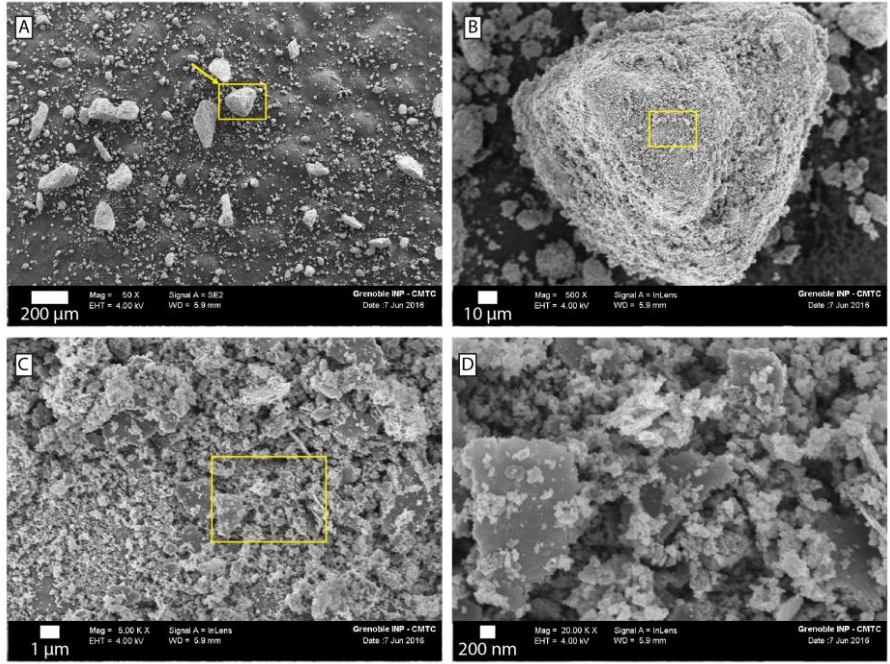

**Fig. 1** - Scanning Electron Microscopy images of pyrrhotite after hand-crushing, 200 μm sieving and colloidal grinding with a planetary grinder PM100 (Retsch). (a) The sample consists of free and agglomerated submicrometer-sized grains. The large clusters are fragile and easily dispersed with a needle under a binocular microscope. (b) Zoom on the agglomerate outlined by the yellow square and arrow in (a). (c, d) The same agglomerate with different magnifications. Grains are submicrometer-sized with a broad range of sizes. Note the presence of platelets, which likely point to re-agglomeration during grinding and/or desiccation.

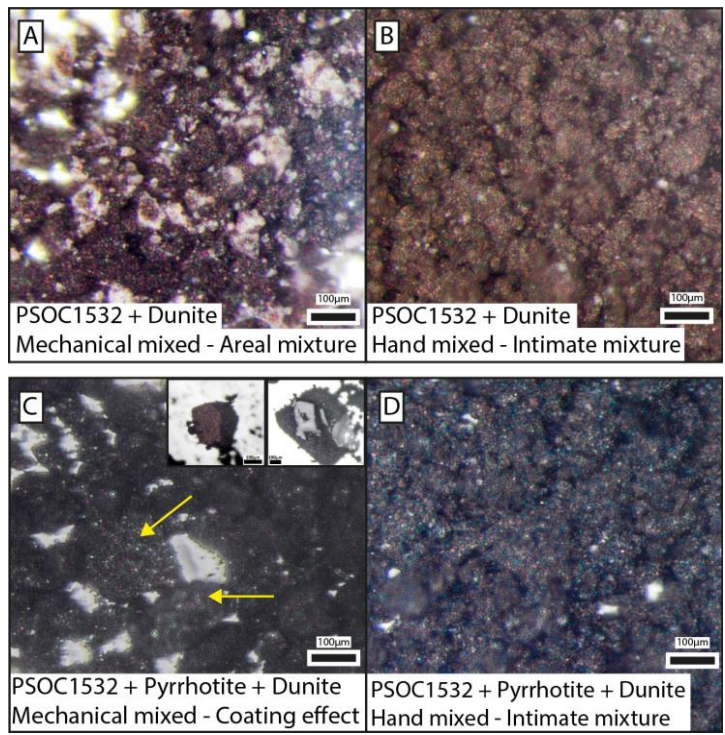

**Fig. 2** - Optical binocular images of different mixtures sieved < 25 μm after a colloidal grinding with a planetary grinder. (A) Areal mixture of PSOC 1532 and dunite mixed with grinder MM200 without milling balls. (B) Intimate mixture of PSOC 1532 and dunite hand-mixed into a mortar with a pestle. (C) PSOC 1532, dunite and pyrrhotite are mixed with the grinder MM200 as sample A. Pyrrhotite coats unbroken aggregates of dunite and organics (yellow arrow). Boxes in the upper right corner show broken aggregates of PSOC 1532 and dunite. (D) Intimate mixture of PSOC 1532, dunite and pyrrhotite, hand mixed into a mortar.

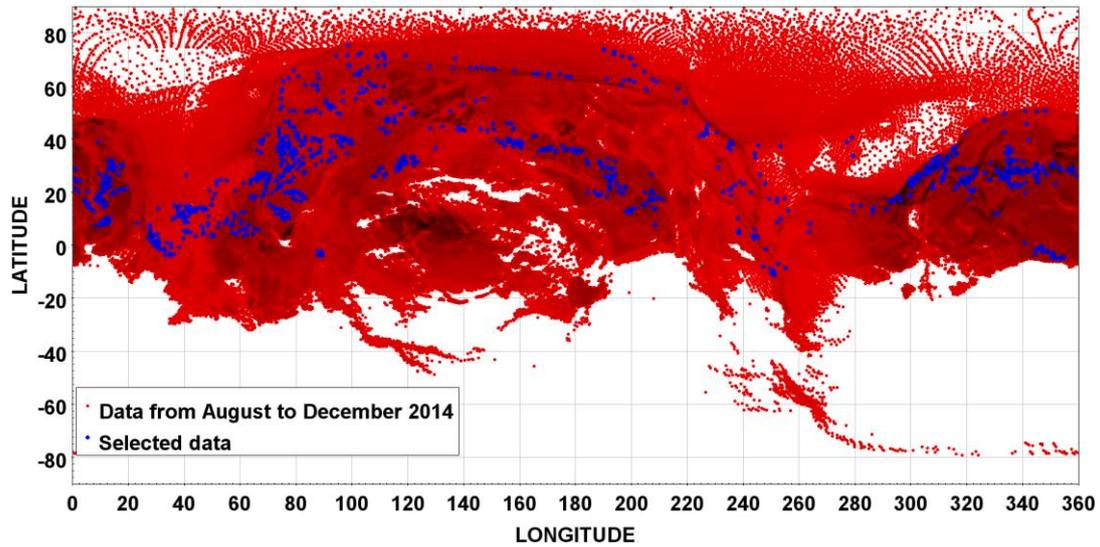

**Fig. 3** - Partial map of the nucleus of 67P (observations from August to December 2014, corresponding to MTP006-MTP011 (MTP means Medium Term Plan). Each coloured point indicates the footprint of a spectrum. Pixels selected to compute the median spectrum of Fig. 4 are displayed in blue. The nucleus surface coverage is partial since the southern hemisphere was not illuminated at this time period.

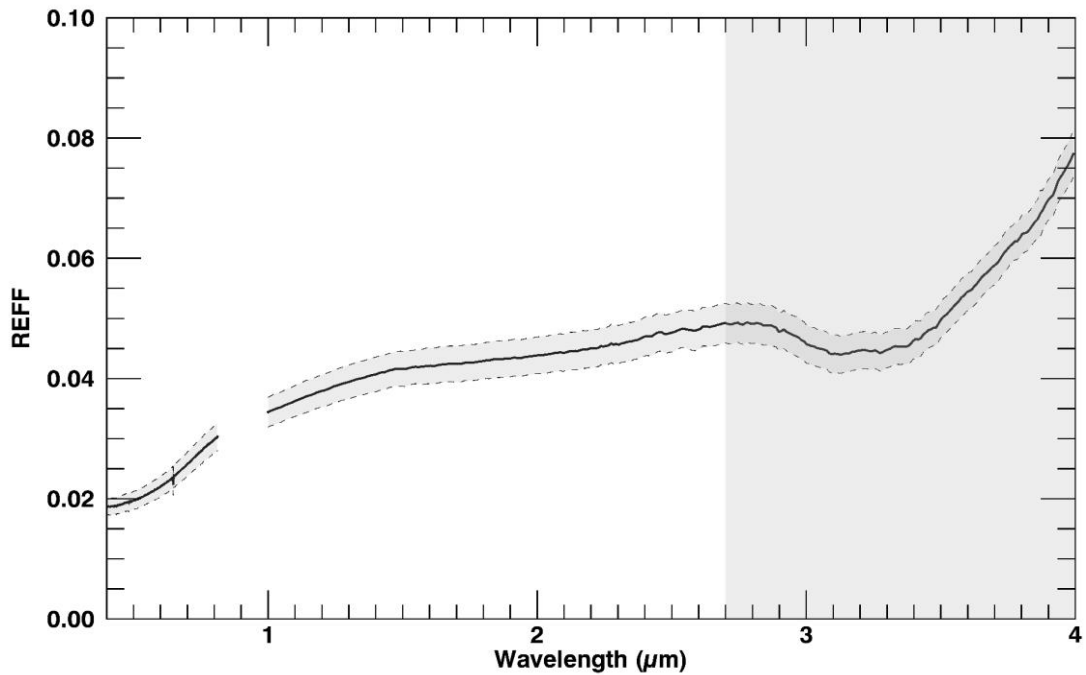

**Fig. 4** - VIRTIS median spectrum of the surface of comet 67P, taken as the reference for this study and plotted in the range of our experimental data (0.4–4.0 μm). At this time of the mission, thermal emission dominates longward of 3.5 μm (not represented). The lack of data between 0.8 μm and 1.0 μm is due to a calibration artefact. Very small variations present at 1.5 μm and 2.5 μm are due to sorting filter junctions. The grey box on the right highlights the spectral range which is not studied because of the presence of the 3.2 μm absorption band, the contribution of the thermal emission and measurement conditions of our mixtures.

### 2.4. VIRTIS Data extraction

The VIRTIS spectrometer (Coradini et al., 2007) has acquired reflectance spectra from two channels: VIRTIS-M, an imaging spectrometer, ranging from 0.25 μm to 5.1 μm with ~2 nm and ~10nm spectral sampling (in the visible and infrared range, respectively) and VIRTIS-H, a point spectrometer ranging from 1.9 μm to 5.1 μm with a ~10 times higher spectral resolution than VIRTIS-M. A selection of spectra acquired by VIRTIS-M in August 2014 (see Fig. 3 and is used in this study. Selecting this short time period minimises activity-related variations of the surface spectral properties (Filacchione et al., 2016; Ciarniello et al., 2016). In addition, since the cometary heliocentric distance was 3.5 AU, the thermal emission relatively was weaker compared to later mission stages. Fig. 4 shows the median of these spectra, calculated from 900 spectra. The selected data were restricted to incidence angles between 0° and 5° and emergence angles between 28° and 32°. For each wavelength, we selected the best observations, then we computed the median spectrum. The error margins in Fig. 4 are given by the standard deviation. In Fig. 3, the footprints of spectra that contribute to the median spectrum are indicated in blue. They are distributed over different regions and therefore encompass the little spectral variability of the part of the surface that was illuminated at that time, mostly in the northern hemisphere. (Ciarniello et al., 2015; Fornasier et al., 2015; Filacchione et al., 2016; Quirico et al., 2016). The map includes acquisitions recorded between August and December 2014 (red dots).

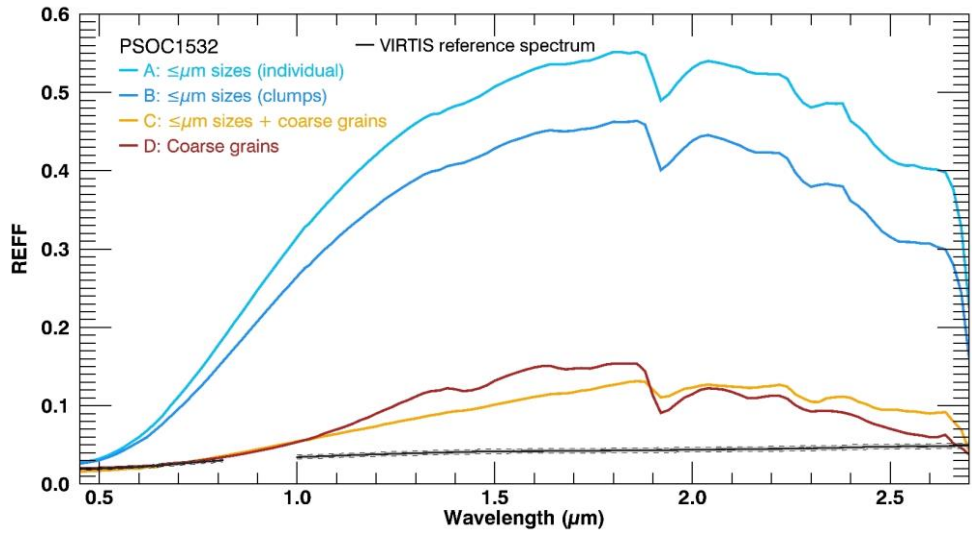

**Fig. 5** - Spectra of PSOC1532 with various grain sizes compared to the VIRTIS spectrum. Spectra A to C correspond to samples after 180 min of colloidal grinding with the planetary grinder. Samples A and B (top) are sieved < 25 μm after the grinding. The sample corresponding to spectrum B contains clumps, unlike sample A. Sample C consists of sub-micrometric particles, clumps and coarse unground grains (not sieved after planetary grinding). The spectrum D corresponds to the original PSOC1532 sample before grinding and sieving.

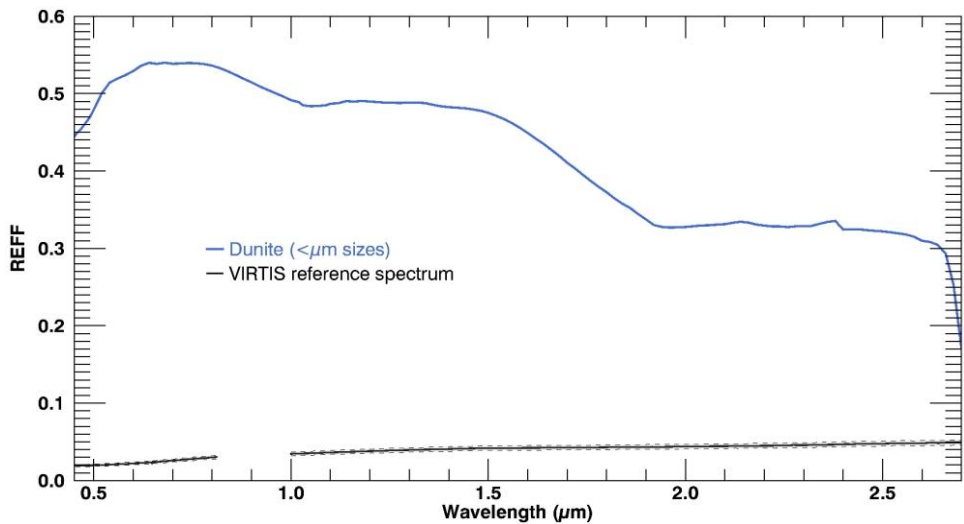

**Fig. 6** - Dunite spectrum compared to VIRTIS reference spectrum. The sample of dunite is sieved < 25 μm after a 180 min colloidal grinding in the planetary grinder.

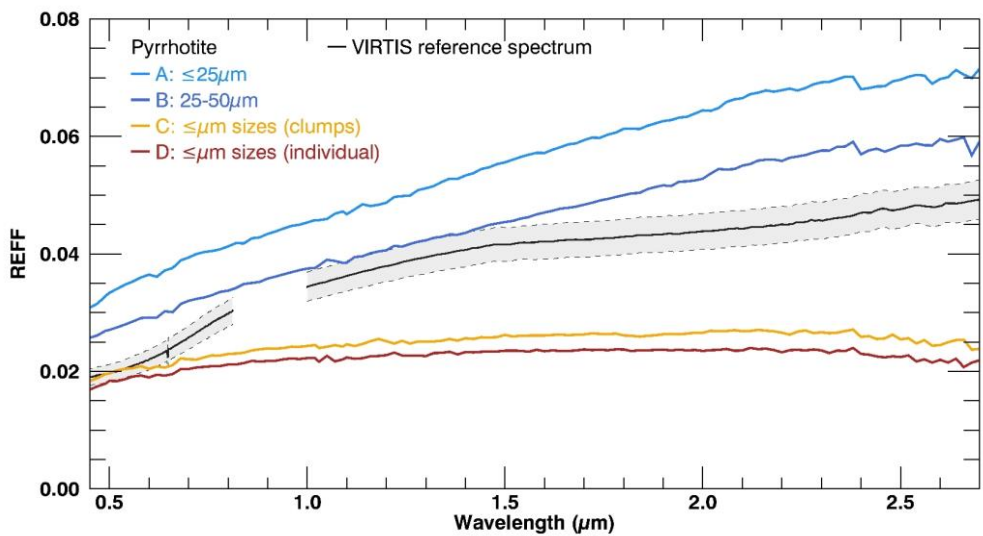

**Fig. 7** - Spectra of ground pyrrhotite. The spectra A and B correspond to hand-ground and sieved samples. The sample C and D underwent 180 min colloidal grinding with the planetary grinder and contain only grains < 25 μm. The sample C contains clumps because it has not been hand mixed after drying unlike the sample of the spectrum D.

The radiance values $I$ measured by VIRTIS are converted to reflectance factors

$$REFF(i, e, g, \lambda) = \frac{I(i, e, g, \lambda)}{F} \frac{1}{\cos(i)}$$

where $F$ is the solar irradiance at 3.5 AU, $i$, $e$ and $g$ are the incidence, emergence and phase angles, respectively, and $\lambda$ is the wavelength. This correction allows one to compare VIRTIS data with data acquired in the laboratory. We observe two different spectral slopes in the 67P nucleus spectra: one in the visible range and the second in the infrared (Capaccioni et al., 2015; Ciarniello et al., 2015; Filacchione et al., 2016). A complex absorption band is visible, from 2.8 µm to 3.5 µm (Capaccioni et al., 2015). This broad feature cannot be attributed to a single compound. According to Quirico et al. (2016), organic -OH, aromatic C-H, $CH_2$ and $CH_3$-groups, $NH_4^+$ ion are plausible species contributing to this feature. The part of the spectrum longward of 3.5 µm is dominated by the thermal emissions from the surface, and even after correction for this thermal emission no clear absorption features have been identified in this range. The present study focuses on the part of the spectrum from 0.4 µm to 2.7 µm.

## 3. Results
### 3.1. Endmembers, the impact of sub-µm grain
#### 3.1.1. PSOC 1532

The coal PSOC 1532 shows a spectrum with a low reflectance in the visible (less than 0.05 at 0.55 µm, see Fig. 5), which increases towards longer wavelengths. The maximum reflectance value is reached around 1.9 µm. A series of overlapping absorption bands is observed between 2.2 and 2.6 µm. These bands are mostly due to combinations of various organic fundamental vibrational modes and to OH stretching vibrations. Note however that OH stretching vibrations have a strong infrared activity but that clays are weakly abundant according to the faint 10 µm band in transmission infrared spectra. Absorption bands due to water and structural OH are visible at 1.9 µm and also at 1.4 µm although less evident. The spectra of the original material (before grinding or sieving) display a low reflectance in the visible (0.02 at 0.55 µm) and a maximum reflectance around 0.15 at 1.9 µm. The latter value is higher by a factor of 3 with respect to the cometary nucleus. When ground to sub-µm grain size, the overall reflectance is increased by a factor of 4. In a previous study (Quirico et al., 2016) the impact of decreasing grain size on the reflectance was determined for grain size down to 25-50 µm, revealing a similar increase in the overall reflectance.

#### 3.1.2. Silicates

The reflectance spectrum of the dunite powder is presented in Fig. 6, where the sample was ground to sub-µm grain size using the planetary grinder. The sample remains relatively bright in the whole spectral range (reflectance above 0.3), with a blue slope starting at ~0.7 µm, which is more pronounced than for a typical dunite. Absorption features are present around 1 and 2 µm. These absorptions are related to the presence of Fe in the silicate (olivine and pyroxene solid solutions) (Burns, 1989). Note that this dunite is natural, and its Fe/(Fe+Mg) is around 20%.

#### 3.1.3. Iron sulphide (pyrrhotite)

The spectra obtained on the pyrrhotite with various grain sizes are presented in Fig. 7. The spectra obtained for the hand-sieved fraction 25–50 µm and < 25 µm show a low reflectance (0.035 and 0.045 at 1 µm respectively) slightly higher than the value that was derived here for the surface of 67P. These two spectra have a strong red slope across the whole spectral range. This spectral slope is typical of metals when there is a contribution of specular reflection to the measured reflectance (Cloutis et al., 2010). The spectrum of the sulphide that was ground to sub-µm size with the planetary grinder has a lower reflectance (0.02 at 1 µm), even lower than that measured for comet 67P. The spectrum is also much flatter than that of sulphide with a larger grain size. This difference is not due to a composition effect (the pyrrhotite structure is preserved) but it is instead a pure grain size effect.

### 3.2. Binary mixtures

In order to assess the spectral behaviour of mixtures with subµm grains, two different experiments were performed. In a first step, grains of dunite were intimately mixed with PSOC 1532 and in a second step, a mixture between dunite and sulphide was prepared. The intimate mixture was obtained by hand-mixing. The silicate + coal mixture (Fig. 8) shows a non-linear behaviour. The spectrum of the mixture is almost identical to the coal spectrum in the region where the coal is strongly absorbing (< 1.5 µm). In the region where the two endmembers are bright (longward of 1.5 µm), the mixture spectra are nearly intermediate between those two. Note that the mixture spectrum is not a good match to the cometary spectrum and a darkening agent is needed in addition to dunite and the organic phase in order to explain the low reflectance of the nucleus from the Vis to the NIR. The spectrum of the intimate mixture between sub-µm sulphide powder (opaque) and a dunite powder (more reflective) is shown in Fig. 9. The mixture has a low reflectance value, its spectral signature being almost identical to that of the sulphide. The behaviour of the mixture is again non-linear, the silicate signature being totally obscured by the presence of the sulphide. The spectrum of the silicate-sulphide binary mixture has a reflectance level lower than the comet, but the red spectral slope is not reproduced.

### 3.3. Ternary mixtures

In Fig. 10, ternary mixture of sub-µm sulphides, coals and silicates are presented. The samples were prepared using a constant sulphide to organics ratio (1/3 vs 2/3 in the relative mass fraction) and increasing fraction of dunite (5, 10, 20, 30wt.%). In these experiments, the C/Fe is roughly solar (10.5 for the mixture against 9.5 for a solar composition, (Lodders, 2003). The C/Si ratio varies according to the proportion of dunite, within a range from 6.3 (5wt.%) to 51.6 (30wt.%) against 8.3 for a solar composition (Lodders, 2003). Given that some of the solar carbon might have been hosted by gaseous species (CO, CN, $CO_2$, HCN) the solar C/Si value is probably an upper limit for cometary C/Si. The spectra of the ternary mixture are similar to those of the sulphide-silicate mixture. VIS slope and albedo approximate the 67P spectrum as measured by VIRTIS. On the contrary, the slope is too flat and albedo too low in the IR. The addition of a significant organic fraction does not help in reproducing the spectral slope typical of cometary dust as observed by VIRTIS. The fraction of silicate has only a marginal impact on the reflectance spectra of the mixture. The various mixture spectra are almost identical with the exception of the visible. Note that this is the region where grain size begins to be of the order of the wavelength and change of scattering regime might occur.

## 4. Discussion
### 4.1. The spectral behaviour of sub-µm particles

Radiative transfer in a particulate medium is a combination of reflection, absorption, and diffraction. When the grain size is much larger than the wavelength, two major scattering regimes are often described. First, a surface scattering regime, when most of the light seen by the observer has undergone one, two or more reflections on grain surfaces. This is the case of strongly absorbing material (sulphides for example). Second, a volume scattering regime, which is typical of materials with low absorption coefficients in the wavelength range of interest, for which most of the light seen by the observer has gone through at least one grain, and often several ones. This is the case of Fe-poor silicates, which are translucent in the VNIR range. These two regimes can be dealt with and modelled according to geometric optics. In general, for the volume scattering regime, the impact of decreasing grain size is to increase the overall reflectance and decrease absorption band depths (Adams and Filice, 1967). This effect has been observed for carbonaceous chondrites (Johnson and Fanale, 1973) or coal materials (Quirico et al., 2016). In the case of surface scattering, the impact of grain size is less clear, since grain shape and orientation can have a first-order impact on the reflectance level. In the case of troilite, the finest grain size studied by Britt et al. (1992) appeared to have the lowest overall reflectance. This is also the case of powdered iron meteorite (Cloutis et al., 2015). When grain size is below the wavelength, geometric optics cannot be used and a change of scattering regime is expected to occur when grain size approaches $\lambda/\pi$ (Hapke, 2012). An isolated particle experiences a strong decrease in its scattering cross section and becomes a perfect absorber. Nevertheless, particles may be not isolated within a regolith, and the behaviour would be different (e.g. leading to a reduced contribution of the diffraction to the scattering efficiency). One would expect then to observe a decrease in the overall reflectance (for both absorbing and non-absorbing materials) when grain size becomes $< \lambda/\pi$ as was observed for silicates in the mid-IR (Mustard and Hays, 1997). These effects are clearly not observed for our sub-µm samples of silicates and coals. Both samples when ground below 0.5 µm still remain generally bright and do not appear to have experienced a change of scattering regime. The deviation from the theoretically expected behaviour is due to the presence of inter-particle interaction. The presence of electromagnetic coupling between grains (grains sticking was often observed under the SEM) might lead to a difference between the physical grain size (< 1µm) and the "optical grain size". In other words, our reflectance spectra of sub-µm silicate and coals very much resemble the VNIR spectra of 10 µm-sized samples. In the case of the sulphides, a different behaviour is observed, which appears to be in better agreement with the theoretical prediction. The reflectance spectrum obtained for the pyrrhotite sample with the smallest grain size has a different slope (less red) and a lower overall reflectance with respect to sample with larger grain size. Such a behaviour can be interpreted by a decrease of the contribution of specular reflection. As shown by Britt and Pieters (1988), the reflectance spectra of metallic surfaces of low roughness measured in the specular or near specular direction show a strong red slope and a larger reflectance, while at non-specular geometries the spectrum is darker and characterised by a flat to blue slope (Britt and Pieters, 1988). Therefore, in the case of the sub-µm sulphide, the grain size (and therefore the surface roughness) is low enough to minimise the contribution of a specular (coherent) reflection by any part of the sample. This "quenching" of the specular reflection is likely to produce the change of spectra observed for our sub-µm sulphide. Contrary to the coal and the dunite, the pyrrhotite is less subject to formation of aggregates, resulting in an optical grain size closer to the actual physical grain size.

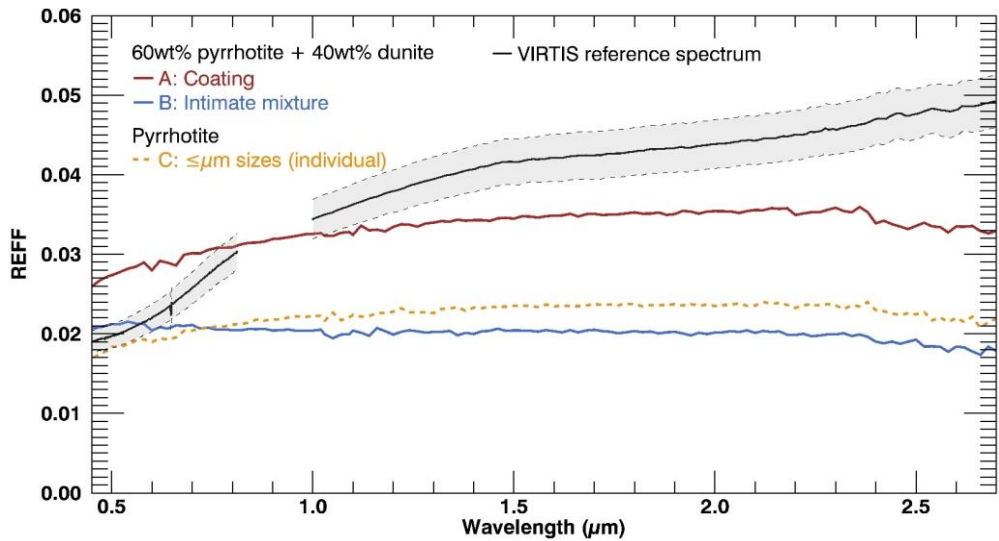

**Fig. 8** - Reflectance spectra of: (A) intimate (see Fig. 2 .B) and (B) areal mixtures (see Fig. 2 .A) of PSOC 1532 and dunite compared to VIRTIS reference spectrum and pure PSOC 1532 spectra. Samples A and C are produced by hand mix while samples B is produced by mechanical mix (preserving clumps). Note that no coating occurs for the areal mixture.

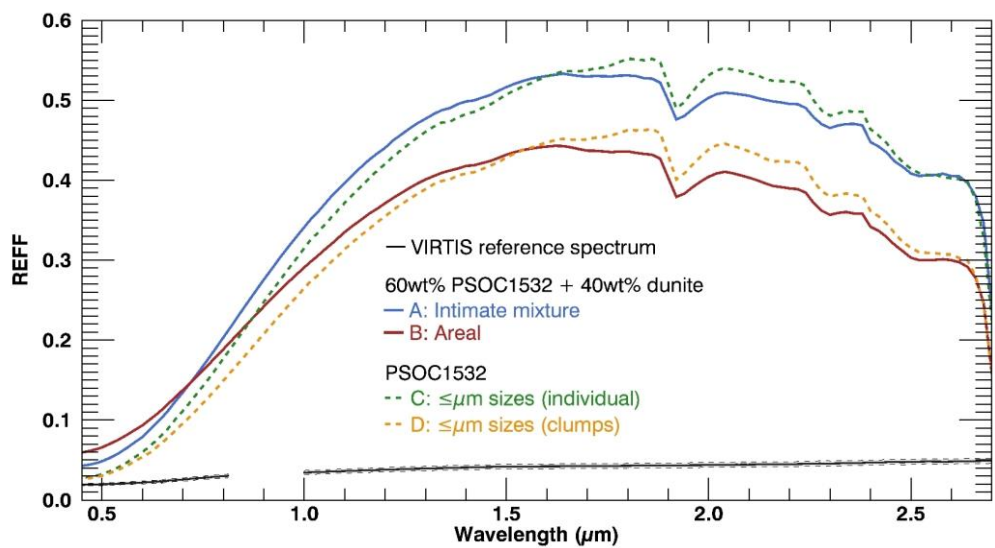

**Fig. 9** - Reflectance spectra of (A) mixture with coating and (B) intimate mixture of 60 wt% pyrrhotite and 40 wt% dunite compared to VIRTIS reference spectrum and a pure sample of pyrrhotite (C). The sample C does not contain clumps and has the same texture as mixture B.

### 4.2. Spectral behaviour of mixtures

The spectral behaviour of a sub-µm grains powders was investigated also for sulphide/silicate, as well as for coal/silicate mixtures (Figs. 8 and 9). The corresponding spectra clearly show a non-linear behaviour, i.e. the reflectance of the mixture is not the weighted average of the endmember spectra. In both cases, the reflectance of the mixture appears to be dominated by the dark material when present. The coal-dunite mixture has a spectrum close to the coal spectra below 1.5 µm. In the wavelength range where both materials are relatively bright (above 1.5 μm), the reflectance of the mixture is intermediate between the two. In the case of the dunite-sulphide mixture, the spectrum of the mixture is almost identical to that of the pyrrhotite. When mixing two endmembers (bright and dark) with similar grain size, the reflectance of the mixture is non-linear and tends to be "biased" toward the dark material (i.e. Pommerol and Schmitt, 2008) but not to the extent observed here for the sulphide-dunite mixture or the coal-dunite mixture (the coal being quite dark in the IR). Previous laboratory work on grain-size > 10 μm has shown that darkening can occur when a bright material is mixed with a darker component (Pommerol and Schmitt, 2008; Yoldi et al., 2015). When the grain sizes of the constituents are similar, darkening is more efficient than when the grain size of the dark material is higher than that of the bright component (for a given volume mixing ratio). In this work, we show that a strong darkening can be obtained when mixing two materials with similar grain size below the micron size. This observation is reinforcing our earlier suggestion that sub-μm sulphides and metals are prime actors in explaining the darkness of cometary surfaces (Capaccioni et al., 2015; Quirico et al., 2016). When mixing a non-icy bright and dark materials with similar grain size $\gg \lambda/\pi$, photons travelling through the media are reflected or scattered by the bright material and reflected or absorbed by the dark one. The overall reflectance of the mixture can therefore be brighter than that of the dark endmember due to reflections and scattering by the bright material. When grain size is below $\lambda/\pi$, we can suppose that reflection and scattering by the bright material become negligible and that the dark material will ultimately absorb any photon travelling through the media. The spectrum of the mixture is therefore only controlled by the absorption properties of the dark material.

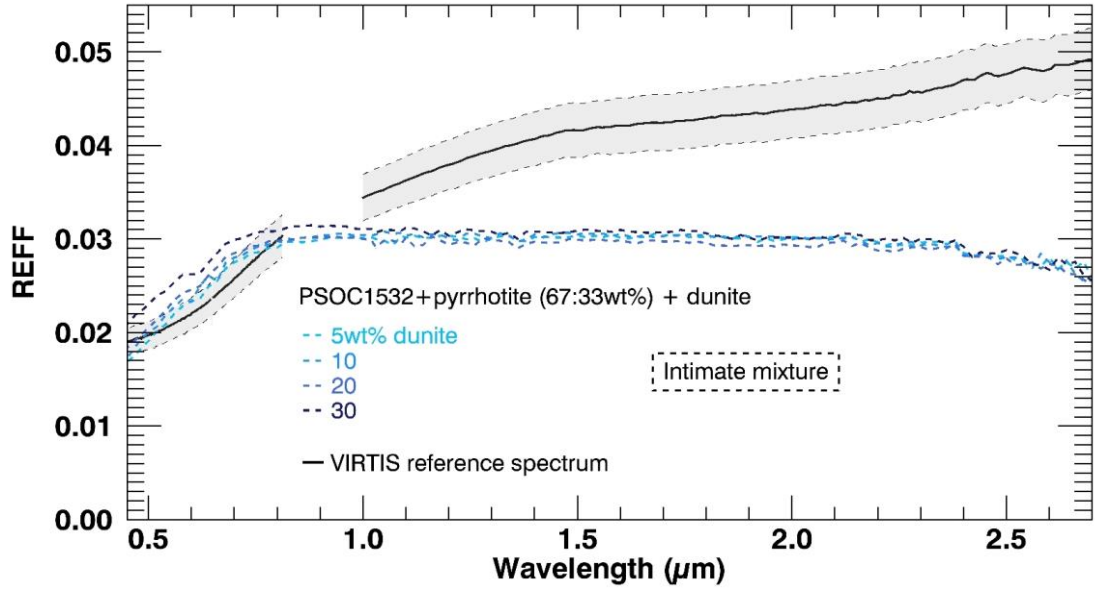

**Fig. 10** - Reflectance spectra of intimate mixtures of PSOC 1532, pyrrhotite and dunite. The coal/pyrrhotite ratio is kept constant. The fraction of dunite varies from 5 wt% to 30 wt% but does not lead to strong variations of the spectral shape.

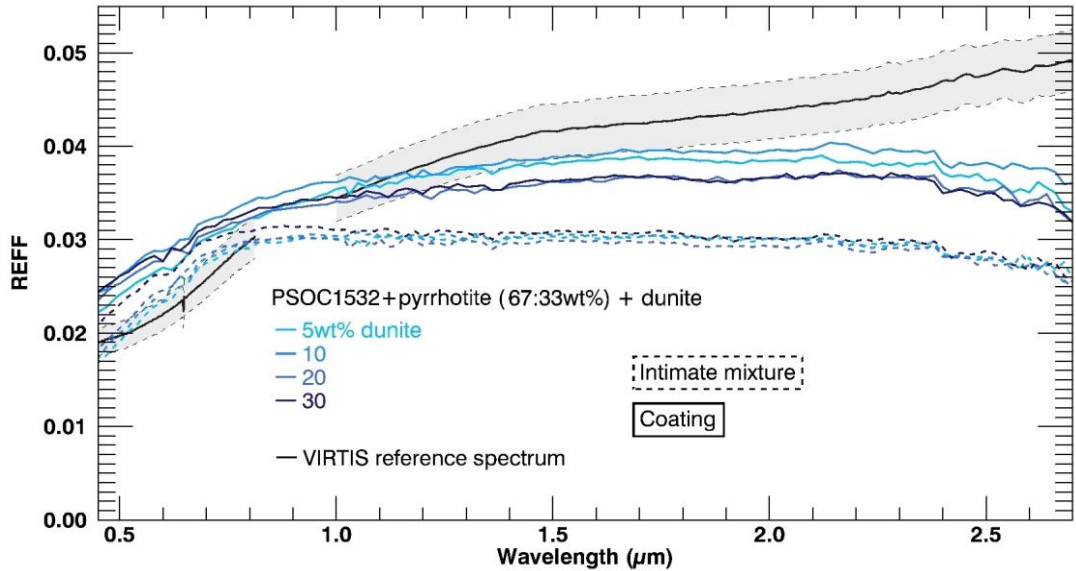

**Fig. 11** - Same figure as Fig. 10 with mechanically mixed ternary mixtures of PSOC 1532, pyrrhotite and dunite. This protocol preserves clumps of organic and silicate which are covered by pyrrhotite (coating mixture, see text). The resulting spectra are similar to the ones of the intimate mixtures in the visible while they are redder and brighter in the infrared range. Various proportions of dunite do not affect strongly the spectral behaviour.

### 4.3. How to reproduce both the spectral slope and the low reflectance level?

A first attempt to reproduce the reflectance spectra of the cometary nucleus was done through ternary mixtures of silicate, coal and sulphide (Fig. 10). The coal/sulphide ratio was kept constant (2/3 coal, 1/3 sulphide in weight) with a C/Fe about solar, and increasing fractions of silicate were added to this mixture (5, 10, 20, 30wt.%). Mixtures were prepared manually by grinding in an agate mortar, which insured intimate mixing at the finest scale (Figs. 1 and 2). The spectra of these mixtures are almost identical to those of the sub-μm sulphide, confirming that intimate mixtures of sub-μm material are strongly dominated by absorbing particles when present. Still, a small impact of the silicate content is observed around 0.7 μm, but this could be due to a change of scattering regime since grain size becomes closer to the wavelength. These ternary intimate mixtures failed to reproduce the 67P's spectrum. A first possibility is that the type of mixing used in our experiments is not adequate, and that not all compounds are intimately mixed, but that large aggregates of a given component are present. The identification of organics-rich grains in 67P (Fray et al., 2016) would suggest some level of heterogeneity in the dust-grain aggregates. In order to produce some level of heterogeneity in our samples, mixtures of sub-µm grains were prepared using a mechanical grinder. Such a preparation protocol was able to preserve clumps formed during the drying of pure phases (especially for silicate and organics). In such a sample, these 100 µm-sized aggregates are covered by pyrrhotite (Fig. 2). If the coating is not complete, the sample has the potential to behave more like an areal mixture with the reflectance of the mixture being a linear combination of the spectral endmembers weighted by the fraction of covered area. This type of mixture has the potential to produce a dark and red spectrum, more similar to 67P's (Fig. 11). Although the spectra of the mixture obtained are not yet a perfect match to the one of the comet, with the mixture reflectance levels too low in the IR, a red spectral slope is present up to 2.3 µm. This slope is likely due to the presence of a few organics-rich aggregates in the sample, which, unlike the sulphide, preserve their spectral slope when ground below the µm. Other alternatives can be proposed to explain the spectral slope of comet 67P surface material. First, a sample that mixes sub-µm grains with other sizes (e.g. ~100-200 µm) might be necessary to retrieve the spectral slope of the VIRTIS spectrum (Moroz et al., 2016; Rousseau et al., 2016; 2017) while being more representative of a comet surface. In such a sample, however, the scattering behaviour is complex. The organics material used in these experiments might not be red enough (not H-rich enough, Moroz et al. (1998)) to impact the spectral slope when intimately mixed grains of organics, sulphides and silicates are prepared (with sub-µm grains). In situ measurements by the COSIMA instrument suggest that the bulk organic compounds found in 67P particles are similar to the insoluble organic matter (IOM) found in meteorite and IDP (the IOM which is relatively similar to our coal) but that it is richer in hydrogen than the IOM (Fray et al., 2016). Another possibility is that our prepared mixture has a too high sulphide/organic ratio, which tends to erase the organic signatures. These experiments were designed in order to have C/Fe ratio close to solar. This value was taken as an upper bound since some of the solar carbon should have been distributed within gaseous species (CO, CN, $CO_2$, HCN) in addition to organic compounds, while most of the iron is expected to have been present in a condensed reduced phase in early Solar System grains (Fe-metal and Fe-sulphides). Still, some of the iron could have been present within the silicates. Further studies will include mixtures with higher organic/sulphide ratios.

## 5. Conclusions

From the obtained results the following conclusions can be drawn:

• A method was developed to produce large samples made of sub-µm individual grains which enables to produce cosmochemically relevant cometary analogues. Therefore, we can study a radiative transfer regime where the physical grain size is below the wavelength. Intimate mixtures of organics (coal), sulphides and silicates were produced in which the typical grain size is a few 100s nanometres.

• When ground below the µm scale, a brightening is observed for moderately absorbing materials (coal and silicate) when compared to grain size > µm, while one would expect a darkening given that the scattering cross section of particles with size smaller than the wavelength should decrease. This suggests that for this type of material the "optical" grain size is still larger than wavelength because of clumping of the particles due to electro-magnetic interactions. In the case of the opaque material (sulphides), when ground below the µm, a darkening is observed as well as a decrease in the spectral slope. This effect is interpreted as due to the reduction of the scattering cross section for very small particles (size < $\lambda/\pi$) and by a decrease of specular reflection contributions to the measured reflectance.

• Sub-µm sulphides can efficiently darken a sample within intimate mixtures of sub-µm materials. This reinforces earlier suggestions (Capaccioni et al., 2015; Quirico et al., 2016) that Ferich opaques (sulphides and (Fe, Ni) metals) are the major contributor to the darkness of 67P and probably comets in general.

• The prepared intimate mixtures are able to reproduce the low reflectance levels observed for comet 67P, but the spectral slope remains too small. A possibility that was explored is that a few percent of organic-rich aggregates may induce this slope. A mechanically prepared mixture where organic-rich aggregates are present (as well as sulphide-rich and dunite-rich aggregates) show a more pronounced spectral slope.

• The peculiar slope of 67P, cometary nuclei and D-type asteroids may be explained by the presence of H-rich organics (redder than the used coal) and by a larger organic/opaque ratio than used in this study.

## Acknowledgments

We thank the following institutions and agencies that support this work: Italian Space Agency (ASI, Italy), Centre National d' Etudes Spatiales (CNES, France), Deutsches Zentrum für Luftund Raumfahrt (DLR - Germany), NASA (USA) Rosetta Program, and Science and Technology Facilities Council (UK). VIRTIS was built by a consortium, which includes Italy, France, and Germany, under the scientific responsibility of the Istituto di Astrofisica e Planetologia Spaziali of INAF, Italy, which also guides the scientific operations. The VIRTIS instrument development has been funded and managed by ASI, with contributions from Observatoire de Meudon financed by CNES, and from DLR. UniversitéGrenoble Alpes (UGA) and CNES are warmly acknowledged for their support to instrumental facilities and activities at IPAG. The Cold Surface Spectroscopy Facility in IPAG is partly funded by. the Europlanet H2020 Research Infrastructure project, which has received funding from the European Union's Horizon 2020 research and innovation programme under grant agreement No 654208. The authors would like to thank Antoine Pommerol and the anonymous reviewer for their constructive comments that substantially improved the manuscript.

## Appendix

**Table 2** - List of the cubes used for compiling the median VIRTIS spectrum. They were acquired at the beginning of the Rosetta mission at the comet during MTP006/STP013 (14 & 15 August 2014).

| Visible | Infrared |
| --- | --- |
| V1_00366679119 | I1_00366679117 |
| V1_00366686318 | I1_00366686316 |
| V1_00366693519 | I1_00366693517 |
| V1_00366700719 | I1_00366700717 |
| V1_00366707918 | I1_00366707916 |
| V1_00366725919 | I1_00366725917 |
| V1_00366729519 | I1_00366729517 |
| V1_00366733119 | I1_00366733117 |
| V1_00366736719 | I1_00366736717 |
| V1_00366740319 | I1_00366740317 |
| V1_00366743919 | I1_00366743917 |
| V1_00366747519 | I1_00366747517 |
| V1_00366751119 | I1_00366751117 |
| V1_00366754719 | I1_00366754717 |
| V1_00366758319 | I1_00366758317 |
| V1_00366765519 | I1_00366765517 |

## References

Adams, J.B., Filice, A.L., 1967. Spectral reflectance 0.4 to 2.0 microns of silicate rock powders. J. Geophys. Res. 72, 5705.

Aléon, J., Engrand, C., Robert, F., Chaussidon, M., 2001. Clues to the origin of interplanetary dust particles from the isotopic study of their hydrogen-bearing phases. Geochim. Cosmochim. Acta 65, 4399–4412. http://www.sciencedirect.com/science/article/pii/S0016703701007207.

Alexander, C.M.O., Fogel, M., Yabuta, H., Cody, G.D., 2007. The origin and evolution of chondrites recorded in the elemental and isotopic compositions of their macromolecular organic matter. Geochim. Cosmochim. Acta 71, 4380–4403.

Bentley, M.S., Schmied, R., Mannel, T., Torkar, K., Jeszenszky, H., Romstedt, J., Levasseur-Regourd, A.-C., Weber, I., Jessberger, E.K., Ehrenfreund, P., Koeberl, C., Havnes, O., 2016. Aggregate dust particles at comet 67P/Churyumov–Gerasimenko. Nature 537 (7618), 73–75. https://doi.org/10.1038/nature19091.

Bockelée-Morvan, D., Rinaldi, G., Érard, S., Leyrat, C., Capaccioni, F., Filacchione, G., Drossart, P., Migliorini, A., Quirico, E., Tozzi, G., Biver, N., Crovisier, J., Arnold, G., Capria, M.-T., Combes, M., Combi, M., de Sanctis, M.-C., Encrenaz, T., Fink, U., Ip, W., Piccioni, G., Schmitt, B., 2017. Comet 67P outbursts and quiescent coma at 1.3 AU from the Sun: dust properties from Rosetta/VIRTIS-H observations. Monthly Notices of the Royal Astronomical Society, Volume 469, Issue Suppl_2, 21 July 2017, Pages S443–S458, https://doi.org/10.1093/mnras/stx1950.

Bonnefoy, N., 2001. Développement d'un spectrophoto-goniométre pour l'étude de la réflectance bidirectionnelle de surfaces géophysiques : application au soufre et perspectives pour le satellite io. UniversitéJoseph Fourier Grenoble 1 Ph.D. thesis. http://www.sudoc.fr/060156244.

Brissaud, O., Schmitt, B., Bonnefoy, N., Douté, S., Rabou, P., Grundy, W., Fily, M., 2004. Spectrogonio radiometer for the study of the bidirectional reflectance and polarization functions of planetary surfaces. 1. Design and tests. Appl. Opt. 43 (9), 1926–1937. http://ao.osa.org/abstract.cfm?URI=ao4391926.

Britt, D., Bell, J., Haack, H., Scott, E., 1992. The reflectance spectrum of troilite. In: Lunar and Planetary Science Conference. Vol. 23 of Lunar and Planetary Science Conference, Marcg, pp. 167–168.

Britt, D.T., Pieters, C.M., 1988. Bidirectional reflectance properties of iron-nickel meteorites. In: Ryder, G. (Ed.), Lunar and Planetary Science Conference Proceedings. Vol. 18 of Lunar and Planetary Science Conference Proceedings, pp. 503–512.

Burns, R.G., 1989. Spectral mineralogy of terrestrial planets –scanning their surfaces remotely. Mineral. Mag. 53 (370), 135–151.

Capaccioni, F., Coradini, A., Filacchione, G., Érard, S., Arnold, G., Drossart, P., De Sanctis, M.C., Bockelée-Morvan, D., Capria, M.T., Tosi, F., Leyrat, C., Schmitt, B., Quirico, E., Cerroni, P., Mennella, V., Raponi, A., Ciarniello, M., Mc Cord, T., Moroz, L., Palomba, E., Ammannito, E., Barucci, M.A., Bellucci, G., Benkhoff, J., Bibring, J.P., Blanco, A., Blecka, M., Carlson, R., Carsenty, U., Colangeli, L., Combes, M., Combi, M., Crovisier, J., Encrenaz, T., Federico, C., Fink, U., Fonti, S., Ip, W.H., Irwin, P., Jaumann, R., Kuehrt, E., Langevin, Y., Magni, G., Mottola, S., Orofino, V., Palumbo, P., Piccioni, G., Schade, U., Taylor, F., Tiphene, D., Tozzi, GBeck, P., Biver, N., Bonal, L., Combe, J.-P., Despan, D., Flamini, E., Fornasier, S., Frigeri, A., Grassi, D., Gudipati, M., Longobardo, A., Markus, K., Merlin, F., Orosei, R., Rinaldi, G., Stephan, K., Cartacci, M., Cicchetti, A., Giuppi, S., Hello, Y., Henry, F., Jacquinod, S., Noschese, R., Peter, G., Politi, R., Reess, J.M., Semery, A., 2015. The organic-rich surface of comet 67P/Churyumov–Gerasimenko as seen by VIRTIS/Rosetta. Science 347 (6220). http://www.sciencemag.org/content/347/6220/aaa0628.abstract.

Ciarniello, M., Capaccioni, F., Filacchione, G., Raponi, A., Tosi, F., Sanctis, D., C.Capria, M., M. T.Érard, S., Bockelée-Morvan, D., Leyrat, C., Arnold, G., Barucci, A., Beck, P., Bellucci, G., Fornasier, S., Longobardo, A., Mottola, S., Palomba, E., Quirico, E., Schmitt, B., 2015. Photometric properties of comet 67P/Churyumov-Gerasimenko from VIRTIS-M onboard Rosetta. A&A. https://doi.org/10.1051/0004-6361/201526307.

Ciarniello, M., Raponi, A., Capaccioni, F., Filacchione, G., Tosi, F., De Sanctis, M. C., Kappel, D., Rousseau, B., Arnold, G., Capria, M. T., Barucci, M. A., Quirico, E., Longobardo, A., Kuehrt, E., Mottola, S., Érard, S.,

Bockelée-Morvan, D., Leyrat, C., Migliorini, A., Zinzi, A., Palomba, E., Schmitt, B., Piccioni, G., Cerroni, P., Ip, W.-H., Rinaldi, G., Salatti, M., 2016. The global surface composition of 67P/Churyumov–Gerasimenko nucleus by Rosetta/VIRTIS. ii) diurnal and seasonal variability. monthly notices of the royal astronomical society. http://mnras.oxfordjournals.org/content/early/2016/12/09/mnras.stw3177.abstract.

Cloutis, E.A., Hardersen, P.S., Bish, D.L., Bailey, D.T., Gaffey, M.J., Craig, M.A., 2010. Reflectance spectra of iron meteorites: implications for spectral identification of their parent bodies. Meteorit Planet. Sci. 45 (2), 304–332. https://doi.org/10.1111/j.1945-5100.2010.01033.x.

Cloutis, E.A., Sanchez, J.A., Reddy, V., Gaffey, M.J., Binzel, R.P., Burbine, T.H., Hardersen, P.S., Hiroi, T., Lucey, P.G., Sunshine, J.M., Tait, K.T., 2015. Olivine-metal mixtures: spectral reflectance properties and application to asteroid reflectance spectra. Icarus 252, 39–82.

Colangeli, L., Lopez-Moreno, J.J., Palumbo, P., Rodriguez, J., Cosi, M., Della Corte, V., Esposito, F., Fulle, M., Herranz, M., Jeronimo, J.M., Lopez-Jimenez, A., Epifani, E.M., Morales, R., Moreno, F., Palomba, E., Rotundi, A., 2007. The grain impact analyser and dust accumulator (GIADA) experiment for the Rosetta mission: design, performances and first results. Space Science. Reviews. 128, 803–821.

Coradini, A., Capaccioni, F., Drossart, P., Arnold, G., Ammannito, E., Angrilli, F., Barucci, A., Bellucci, G., Benkhoff, J., Bianchini, G., Bibring, J.P., Blecka, M., Bockelée-Morvan, D., Capria, M.T., Carlson, R., Carsenty, U., Cerroni, P., Colangeli, L., Combes, M., Combi, M., Crovisier, J., De Sanctis, M.C., Encrenaz, E.T., Érard, S., Federico, C., Filacchione, G., Fink, U., Fonti, S., Formisano, V., Ip, W.H., Jaumann, R., Kuehrt, E., Langevin, Y., Magni, G., Mccord, T., Mennella, V., Mottola, S., Neukum, G., Palumbo, P., Piccioni, G., Rauer, H., Saggin, B., Schmitt, B., Tiphene, D., Tozzi, G., 2007. VIRTIS: an imaging spectrometer for the rosetta mission. Space Sci. Rev. 128 (1), 529–559. https://doi.org/10.1007/s1121400691275.

Crovisier, J., Leech, K., Bockelée-Morvan, D., Brooke, T.Y., Hanner, M.S., Altieri, B., Keller, H.U., Lellouch, E., 1997. 1997 the spectrum of comet Hale-Bopp (c/1995 o1) observed with the Infrared Space Observatory at 2.9 astronomical units from the Sun. Science 275 (5308), 1904. http://science.sciencemag.org/content/275/5308/1904.abstract.

Della Corte, V., Rotundi, A., Accolla, M., Sordini, R., Palumbo, P., Colangeli, L., Lopez–Moreno, J.J., Rodriguez, J., Rietmeijer, F.J.M., Ferrari, M., Lucarelli, F., Mazzotta Epifani, E., Ivanovski, S., Aronica, A., Cosi, M., Bussoletti, E., Crifo, J.F., Esposito, F., Fulle, M., Green, S.F., Gruen, E., Herranz, M.L., Jeronimo, J.M., Lamy, P., Lopez Jimenez, A., McDonnell, J.A.M., Mennella, V., Molina, A., Morales, R., Moreno, F., Palomba, E., Perrin, J.M., Rodrigo, R., Weissman, P., Zakharov, V., Zarnecki, J.C., 2014. GIADA: its status after the Rosetta cruise phase and on-ground activity in support of the encounter with comet 67P/Churyumov-Gerasimenko. Journal of Astronomical Instrumentation. 3. 1350011–110.

Della Corte, V., Rotundi, A., Fulle, M., Ivanovski, S., Green, S.F., Rietmeijer, F.J.M., Colangeli, L., Palumbo, P., Sordini, R., Ferrari, M., Accolla, M., Zakharov, V., Mazzotta Epifani, E., Weissman, P., Gruen, E., Lopez-Moreno, J.J., Rodriguez, J., Bussoletti, E., Crifo, J.F., Esposito, F., Lamy, P.L., McDonnell, J.A.M., Mennella, V., Molina, A., Morales, R., Moreno, F., Palomba, E., Perrin, J.M., Rodrigo, R., Zarnecki, J.C., Cosi, M., Giovane, F., Gustafson, B., Ortiz, J.L., Jeronimo, J.M., Leese, M.R., Herranz, M., Liuzzi, V., Lopez-Jimenez, A.C., 2016. 67P/C-G inner coma dust properties from 2.2 AU inbound to 2.0 AU outbound to the Sun. Mon. Not. R. Astron. Soc. 462 (Suppl 1), S210-S219. http://mnras.oxfordjournals.org/content/462/Suppl-1/S210.abstract.

Dobrica, E., Brearley, A., 2011. Crystalline silicates in comet 81P/Wild 2 from the Stardust track 81. Meteorit. Planet. Sci. Suppl. 74, 5413.

Dobrica, E., Engrand, C., Leroux, H., Rouzaud, J.-N., Duprat, J., 2012. Transmission electron microscopy of concordia ultracarbonaceous Antarctic micrometeorites (UCAMMs): Mineralogical properties. Geochim. Cosmochim. Acta 76, 68–82. http://www.sciencedirect.com/science/article/pii/S0016703711006120.

Dobrica, E., Engrand, C., Quirico, E., Montagnac, G., Duprat, J., 2011. Raman characterization of carbonaceous matter in concordia Antarctic micrometeorites. Meteorit. Planet. Sci. 46 (9), 1363–1375. https://doi.org/10.1111/j.1945-5100.2011.01235.x.

Engrand, C., Duprat, J., Dartois, E., Benzerara, K., Leroux, H., Baklouti, D., Bardyn, A., Briois, C., Cottin, H., Fischer, H., Fray, N., Godard, M., Hilchenbach, M., Langevin, Y., Paquette, J., Rynö, J., Schulz, R., Silén, J., Stenzel, O., Thirkell, L., 2016. Variations in cometary dust composition from Giotto to Rosetta, clues to their formation mechanisms. Mon. Not. R. Astron. Soc. 462 (Suppl. 1), 323–330. https://doi.org/10.1093/mnras/stw2844.

Filacchione, G., Capaccioni, F., Ciarniello, M., Raponi, A., Tosi, F., De Sanctis, M.-C., Érard, S., Bockelée-Morvan, D., Leyrat, C., Arnold, G., Schmitt, B., Quirico, E., Piccioni, G., Migliorini, A., Capria, M., Palomba, E., Cerroni, P., Longobardo, A., Barucci, A., Fornasier, S., Carlson, R., Jaumann, R., Stephan, K., Moroz, L.V., Kappel, D., Rousseau, B., Fonti, S., Mancarella, F., Despan, D., Faure, M., 2016. The global surface composition of 67P/CG nucleus by Rosetta/VIRTIS. (i) Prelanding mission phase. Icarus 274, 334–349. http://www.sciencedirect.com/science/article/pii/S0019103516001342.

Fomenkova, M.N., Chang, S., Mukhin, L.M., 1994. Carbonaceous components in the comet Halley dust. Geochimica et Cosmochimica Acta 58 (20), 4503–4512.

Fornasier, S., Hasselmann, P.H., Barucci, M.A., Feller, C., Besse, S., Leyrat, C., Lara, L., Gutierrez, P.J., Oklay, N., Tubiana, C., Scholten, F., Sierks, H., Barbieri, C., Lamy, P.L., Rodrigo, R., Koschny, D., Rickman, H., Keller, H.U., Agarwal, J., A'Hearn, M.F., Bertaux, J.-L., Bertini, I., Cremonese, G., Da Deppo, V., Davidsson, B., Debei, S., De Cecco, M., Fulle, M., Groussin, O., Güttler, C., Hviid, S.F., Ip, W., Jorda, L., Knollenberg, J., Kovacs, G., Kramm, R., Kührt, E., Küppers, M., La Forgia, F., Lazzarin, M., Lopez Moreno, J.J., Marzari, F., Matz, K.-D., Michalik, H., Moreno, F., Mottola, S., Naletto, G., Pajola, M., Pommerol, A., Preusker, F., Shi, X., Snodgrass, C., Thomas, N., Vincent, J.-B., 2015. Spectrophotometric properties of the nucleus of comet 67P/Churyumov-Gerasimenko from the OSIRIS instrument onboard the Rosetta spacecraft. A&A 583. https://doi.org/10.1051/0004-6361/201525901.

Fray, N., Bardyn, A., Cottin, H., Altwegg, K., Baklouti, D., Briois, C., Colangeli, L., Engrand, C., Fischer, H., Glasmachers, A., Grün, E., Haerendel, G., Henkel, H., Höfner, H., Hornung, K., Jessberger, E.K., Koch, A., Krüger, H., Langevin, Y., Lehto, H., Lehto, K., Le Roy, L., Merouane, S., Modica, P., Orthous-Daunay, F.-R., Paquette, J., Raulin, F., Rynö, J., Schulz, R., Silén, J., Siljeström, S., Steiger, W., Stenzel, O., Stephan, T., Thirkell, L., Thomas, R., Torkar, K., Varmuza, K., Wanczek, K.-P., Zaprudin, B., Kissel, J., Hilchenbach, M., Sep. 2016. High-molecular-weight organic matter in the particles of comet 67P/Churyumov-Gerasimenko. Nature advance online publication. https://doi.org/10.1038/nature19320.

Fulle, M., Della Corte, V., Rotundi, A., Rietmeijer, F.J.M., Green, S.F., Weissman, P., Accolla, M., Colangeli, L., Ferrari, M., Ivanovski, S., Lopez-Moreno, J.J., Epifani, E.M., Morales, R., Ortiz, J.L., Palomba, E., Palumbo, P., Rodriguez, J., Sordini, R., Zakharov, V., 2016a. Comet 67P/Churyumov-Gerasimenko preserved the pebbles that formed planetesimals. Mon. Not. R. Astron. Soc. 462 (Suppl 1), S132–S137. http://mnras.oxfordjournals.org/content/462/Suppl-1/S132.abstract.

Fulle, M., Della Corte, V., Rotundi, A., Weissman, P., Juhasz, A., Szego, K., Sordini, R., Ferrari, M., Ivanovski, S., Lucarelli, F., Accolla, M., Merouane, S., Zakharov, V., Mazzotta Epifani, E., López-Moreno, J.J., Rodríguez, J., Colangeli, L., Palumbo, P., Grün, E., Hilchenbach, M., Bussoletti, E., Esposito, F., Green, S.F., Lamy, P.L., McDonnell, J.A.M., Mennella, V., Molina, A., Morales, R., Moreno, F., Ortiz, J.L., Palomba, E., Rodrigo, R., Zarnecki, J.C., Cosi, M., Giovane, F., Gustafson, B., Herranz, M.L., Jerónimo, J.M., Leese, M.R., López-Jiménez, A.C., Altobelli, N., 2015. Density and charge of pristine fluffy particles from comet 67p/churyumov-gerasimenko. Astrophys. J. Lett. 802, L12.

Fulle, M., Marzari, F., Corte, V.D., Fornasier, S., Sierks, H., Rotundi, A., Barbieri, C., Lamy, P.L., Rodrigo, R., Koschny, D., Rickman, H., Keller, H.U., López-Moreno, J.J., Accolla, M., Agarwal, J., A'Hearn, M.F., Altobelli, N., Barucci, M.A., Bertaux, J.L., Bertini, I., Bodewits, D., Bussoletti, E., Colangeli, L., Cosi, M., Cremonese, G., Crifo, J.-F., Deppo, V.D., Davidsson, B., Debei, S., Cecco, M.D., Esposito, F., Ferrari, M., Giovane, F., Gustafson, B., Green, S.F., Groussin, O., Grün, E., Gutierrez, P., Güttler, C., Herranz, M.L., Hviid, S.F., Ip, W., Ivanovski, S.L., Jerónimo, J.M., Jorda, L., Knollenberg, J., Kramm, R., Kührt, E., Küppers, M., Lara, L., Lazzarin, M., Leese, M.R., López-Jiménez, A.C., Lucarelli, F., Epifani, E.M., McDonnell, J.A.M., Mennella, V., Molina, A., Morales, R., Moreno, F., Mottola, S., Naletto, G., Oklay, N., Ortiz, J.L., Palomba, E., Palumbo, P., Perrin, J.-M., Rietmeijer, F.J.M., Rodríguez, J., Sordini, R., Thomas, N., Tubiana, C., Vincent, J.-B., Weissman, P., Wenzel, K.-P., Zakharov, V., Zarnecki, J.C., 2016b. Evolution of the dust size distribution of comet 67P/Churyumov-Gerasimenko from 2.2 AU to perihelion. Astrophys. J. 821 (1), 19. http://stacks.iop.org/0004-637X/821/i=1/a=19.

Hapke, B., 2012. Theory of Reflectance and Emittance Spectroscopy. Cambridge University Press. www.cambridge.org/9780521883498.

Harker, D.E., Woodward, C.E., Wooden, D.H., 2005. The dust grains from 9P/Tempel 1 before and after the encounter with deep impact. Science 310 (5746), 278, http://science.sciencemag.org/content/310/5746/278.abstract.

Henkel, H., Höfner, H., Kissel, J., Koch, A., Apr. 2003. COSIMA mass spectrometer for the Rosetta mission. In: EGS AGU EUG Joint Assembly, p. 10075. http://adsabs.harvard.edu/abs/20 03EAEJA....10 075H.

Hérique, A., Kofman, W., Beck, P., Bonal, L., Buttarazzi, I., Heggy, E., Lasue, J., Levasseur-Regourd, A.C., Quirico, E., Zine, S., 2016. Cosmochemical implications of CONSERT permittivity characterization of 67P/CG. Mon. Not. R. Astron. Soc. 462, S516–S532. Johnson, T.V., Fanale, F.P., 1973. Optical properties of carbonaceous chondrites and their relationship to asteroids. J. Geophys. Res. 78 (35), 8507–8518. https://doi.org/10.1029/JB078i035p08507.

Keller, L.P., Bajt, S., Baratta, G.A., Borg, J., Bradley, J.P., Brownlee, D.E., Busemann, H., Brucato, J.R., Burchell, M., Colangeli, L., d'Hendecourt, L., Djouadi, Z., Ferrini, G., Flynn, G., Franchi, I.A., Fries, M., Grady, M.M., Graham, G.A., Grossemy, F., Kearsley, A., Matrajt, G., Nakamura-Messenger, K., Mennella, V., Nittler, L., Palumbo, M.E., Stadermann, F.J., Tsou, P., Rotundi, A., Sandford, S.A., Snead, C., Steele, A., Wooden, D., Zolensky, M., 2006. Infrared spectroscopy of comet 81P/Wild 2 samples returned by Stardust. Science 314 (5806), 1728–1731. http://www.jstor.org/stable/20035030.

Kolokolova, L., Hanner, M.S., Levasseur-Regourd, A.-C., Gustafson, B., S., A., 2004. Physical Properties of Cometary Dust from Light Scattering and Thermal Emission. The University of Arizona Press. 577–604

Langevin, Y., Hilchenbach, M., Ligier, N., Merouane, S., Hornung, K., Engrand, C., Schulz, R., Kissel, J., Rynö, J., Eng, P., 2016. Typology of dust particles collected by the COSIMA mass spectrometer in the inner coma of 67P/Churyumov Gerasimenko. Icarus 271, 76–97. http://www.sciencedirect.com/science/article/pii/S0019103516000403.

Leroux, H., Cuvillier, P., Zanda, B., Hewins, R.H., 2015. Gems-like material in the matrix of the paris meteorite and the early stages of alteration of CM chondrites. Geochim. Cosmochim. Acta 170, 247–265.

Lisse, C.M., VanCleve, J., Adams, A.C., A'Hearn, M.F., Fernández, Y.R., Farnham, T.L., Armus, L., Grillmair, C.J., Ingalls, J., Belton, M.J.S., Groussin, O., McFadden, L.A., Meech, K.J., Schultz, P.H., Clark, B.C., Feaga, L.M., Sunshine, J.M., 2006. Spitzer spectral observations of the deep impact ejecta. Science 313 (5787), 635–640. http://science.sciencemag.org/content/313/5787/635.abstract.

Lodders, K., 2003. Solar System abundances and condensation temperatures of the elements. Astrophys. J. 591, 1220–1247. http://adsabs.harvard.edu/abs/2003ApJ...591.1220L.

Mannel, T., Bentley, M.S., Schmied, R., Jeszenszky, H., Levasseur-Regourd, A.C., Romstedt, J., Torkar, K., 2016. Fractal cometary dust –a window into the early Solar System. Mon. Not. R. Astron. Soc. 462, S304–S311.

Merouane, S., Zaprudin, B., Stenzel, O., Langevin, Y., Altobelli, N., Della Corte, V., Fischer, H., Fulle, M., Hornung, K., Silén, J., Ligier, N., Rotundi, A., Ryno, J., Schulz, R., Hilchenbach, M., Kissel, J., the COSIMA team, 2016. Dust particle flux and size distribution in the coma of 67P/Churyumov-Gerasimenko measured in situ by the COSIMA instrument on board Rosetta. A&A 596. https://doi.org/10.1051/0004-6361/201527958.

Moroz, L., Arnold, G., Korochantsev, A., Wasch, R., 1998. Natural solid bitumens as possible analogs for cometary and asteroid organics. Icarus 134 (2), 253–268. http://www.sciencedirect.com/science/article/pii/S0019103598959553.

Moroz, L.V., Markus, K., Arnold, G., Henckel, D., Kappel, D., Schade, U., Rousseau, B., Quirico, E., Schmitt, B., Capaccioni, F., Bockelée-Morvan, D., Filacchione, G., Érard, S., Leyrat, C., Team, V., Oct. 2016. Reflectance spectroscopy of natural organic solids, iron sulfides and their mixtures as refractory analogues for Rosetta/VIRTIS' surface composition analysis of 67P/cg. AAS/Division for Planetary Sciences Meeting Abstracts. Vol. 48 of AAS/Division for Planetary Sciences Meeting Abstracts. p. 116.21

Mustard, J.F., Hays, J.E., 1997. Effects of hyperfine particles on reflectance spectra from 0.3 to 25 μm. Icarus 125, 145–163.

Pommerol, A., Schmitt, B., 2008. Strength of the H2O near-infrared absorption bands in hydrated minerals: effects of particle size and correlation with albedo. J. Geophys. Res. (Planets) 113 (E12), E10 0 09.

Quirico, E., Moroz, L., Schmitt, B., Arnold, G., Faure, M., Beck, P., Bonal, L., Ciarniello, M., Capaccioni, F., Filacchione, G., Érard, S., Leyrat, C., BockeléeMorvan, D., Zinzi, A., Palomba, E., Drossart, P., Tosi, F., Capria, M., De Sanctis, M., Raponi, A., Fonti, S., Mancarella, F., Orofino, V., Barucci, A., Blecka, M., Carlson, R., Despan, D., Faure, A., Fornasier, S., Gudipati, M., Longobardo, A., Markus, K., Mennella, V., Merlin, F., Piccioni, G., Rousseau, B., Taylor, F., 2016. Refractory and semi-volatile organics at the surface of comet 67P/Churyumov-Gerasimenko: insights from the VIRTIS/Rosetta imaging spectrometer. Icarus 272, 32–47. http://www.sciencedirect.com/science/article/pii/S001910351600097X.

Riedler, W., Torkar, K., Jeszenszky, H., Romstedt, J., Alleyne, H.S.C., Arends, H., Barth, W., Biezen, J.V.D., Butler, B., Ehrenfreund, P., Fehringer, M., Fremuth, G., Gavira, J., Havnes, O., Jessberger, E.K., Kassing, R., Klöck, W., Koeberl, C., Levasseur-Regourd, A.C., Maurette, M., Rüdenauer, F., Schmidt, R., Stangl, G., Steller, M., Weber, I., 2007. MIDAS the micro-imaging dust analysis system for the Rosetta mission. Space Sci. Rev. 128, 869–904. http://adsabs.harvard.edu/abs/2007SSRv..128..869R.

Rotundi, A., Sierks, H., Della Corte, V., Fulle, M., Gutierrez, P.J., Lara, L., Barbieri, C., Lamy, P.L., Rodrigo, R., Koschny, D., Rickman, H., Keller, H.U., López–Moreno, J.J., Accolla, M., Agarwal, J., A'Hearn, M.F., Altobelli, N., Angrilli, F., Barucci, M.A., Bertaux, J.-L., Bertini, I., Bodewits, D., Bussoletti, E., Colangeli, L., Cosi, M., Cremonese, G., Crifo, J.-F., Da Deppo, V., Davidsson, B., Debei, S., De Cecco, M., Esposito, F., Ferrari, M., Fornasier, S., Giovane, F., Gustafson, B., Green, S.F., Groussin, O., Grün, E., Güttler, C., Herranz, M.L., Hviid, S.F., Ip, W., Ivanovski, S., Jerónimo, J.M., Jorda, L., Knollenberg, J., Kramm, R., Kührt, E., Küppers, M., Lazzarin, M., Leese, M.R., López-Jiménez, A.C., Lucarelli, F., Lowry, S.C., Marzari, F., Epifani, E.M., McDonnell, J.A.M., Mennella, V., Michalik, H., Molina, A., Morales, R., Moreno, F., Mottola, S., Naletto, G., Oklay, N., Ortiz, J.L., Palomba, E., Palumbo, P., Perrin, J.-M., Rodríguez, J., Sabau, L., Snodgrass, C., Sordini, R., Thomas, N., Tubiana, C., Vincent, J.-B., Weissman, P., Wenzel, K.-P., Zakharov, V., Zarnecki, J.C., 2015. Dust measurements in the coma of comet 67P/Churyumov-Gerasimenko inbound to the Sun. Science 347 (6220).

Rousseau, B., Beck, P., Érard, S., Quirico, E., Schmitt, B., Bonal, L., Montes-Hernandez, G., Capaccioni, F., Filacchione, G., Bockelée-Morvan, D., Leyrat, C., Arnold, G., Ciarniello, M., Raponi, A., Longonbardo, A., Moroz, L., Palomba, E., Tosi, F., the VIRTIS science team, 2017. Reproducing the vnir spectra of 67P/CG in the laboratory: clues to the composition of surface dust. In: Asteroids, Comets, Meteors, p. 170.

Rousseau, B., Érard, S., Beck, P., Quirico, E., Schmitt, B., Bonal, L., Montes-Hernandez, G., Moroz, L., Kappel, D., Markus, K., Arnold, G., Ciarniello, M., Raponi, A., Longobardo, A., Capaccioni, F., Filacchione, G., Bockelee-Morvan, D., Leyrat, C., Team, R.V., Oct. 2016. Sulfides and refractory organic matter at the surface of 67P/Churyumov-Gerasimenko: evidence from VIRTIS data and laboratory measurements. AAS/Division for Planetary Sciences Meeting Abstracts. Vol. 48 of AAS/Division for Planetary Sciences Meeting Abstracts. p. 211.08

Sugita, S., Ootsubo, T., Kadono, T., Honda, M., Sako, S., Miyata, T., Sakon, I., Yamashita, T., Kawakita, H., Fujiwara, H., Fujiyoshi, T., Takato, N., Fuse, T., Watanabe, J., Furusho, R., Hasegawa, S., Kasuga, T., Sekiguchi, T., Kinoshita, D., Meech, K.J., Wooden, D.H., Ip, W.H., A'Hearn, M.F., 2005. Subaru telescope observations of deep impact. Science 310 (5746), 274. http://science.sciencemag.org/content/310/5746/274.abstract.

Vernazza, P., Beck, P., November 2016. Composition of Solar System small bodies. arxiv e-prints.

Yoldi, Z., Pommerol, A., Jost, B., Poch, O., Gouman, J., Thomas, N., 2015. VIS-NIR reflectance of water ice/ regolith analogue mixtures and implications for the detectability of ice mixed within planetary regoliths. Geophys. Res. Lett.. https://doi.org/10.1002/2015GL064780.

Zolensky, M.E., Zega, T.J., Yano, H., Wirick, S., Westphal, A.J., Weisberg, M.K., Weber, I., Warren, J.L., Velbel, M.A., Tsuchi yama, A., Tsou, P., Toppani, A., Tomioka, N., Tomeoka, K., Teslich, N., Taheri, M., Susini, J., Stroud, R., Stephan, T., Stadermann, F.J., Snead, C.J., Simon, S.B., Simionovici, A., See, T.H., Robert, F., Rietmeijer, F.J.M., Rao, W., Perronnet, M.C., Papanastassiou, D.A., Okudaira, K., Ohsumi, K., Ohnishi, I., Nakamura-Messenger, K., Nakamura, T., Mostefaoui, S., Mikouchi, T., Meibom, A., Matrajt, G., Marcus, M.A., Leroux, H., Lemelle, L., Le, L., Lanzirotti, A., Langenhorst, F., Krot, A.N., Keller, L.P., Kearsley, A.T., Joswiak, D., Jacob, D., Ishii, H., Harvey, R., Hagiya, K., Grossman, L., Grossman, J.N., Graham, G.A., Gounelle, M., Gillet, P., Genge, M.J., Flynn, G., Ferroir, T., Fallon, S., Ebel, D.S., Dai, Z.R., Cordier, P., Clark, B., Chi, M., Butterworth, A.L., Brownlee, D.E., Bridges, J.C., Brennan, S., Brearley, A., Bradley, J.P., Bleuet, P., Bland, P.A., Bastien, R., 2006. Mineralogy and petrology of comet 81P/Wild 2 nucleus samples. Science 314 (5806), 1735–1739. http://science.sciencemag.org/content/314/5806/1735.abstract.